\documentclass[italian]{article}

\usepackage{graphicx}

\title{\bf Uranio impoverito: perch\'e?}

\author{Germano D'Abramo\\
{\small Istituto di Astrofisica Spaziale e Fisica Cosmica,}\\
{\small Area di Ricerca CNR Tor Vergata, Roma, Italy}\\
{\small E--mail: {\tt dabramo@rm.iasf.cnr.it}}}

\date{\small Gennaio-Marzo 2002}

\begin{document}

\maketitle

\section{Introduzione}

Per gran parte degli anni '90 il mondo ha assistito a numerosi conflitti
che hanno pi\`u o meno direttamente coinvolto anche il nostro paese:
dapprima l'invasione del Kuwait da parte dell'Iraq, poi le guerre
intestine nell'area Balcanica (ex--Jugoslavia ed Albania--Kosovo).  In
entrambi i casi le forze armate della cosiddetta Alleanza Atlantica sono
intervenute, e lo hanno fatto con il dispiegamento di mezzi (militari)
proprio delle nazioni forti. \`E stato in queste tristi occasioni che sono
venuto a conoscenza per la prima volta delle armi all'uranio impoverito
(UI o DU, dall'inglese ``depleted uranium'').

Inoltre, alla fine dell'anno 2000 e nei primi mesi del 2001 i mezzi di
comunicazione, soprattutto quelli italiani, hanno riservato ampio spazio
ai proiettili all'UI, ritenuti responsabili di aver procurato gravi
malattie ai soldati e alla popolazione civile durante le operazioni
militari in Albania--Kosovo.

Tralasciando le pur gravi implicazioni di carattere medico e
civile\footnote{ Quest'ultima problematica \`e ovviamente importante e
delicata.}, prioritaria \`e qui la curiosit\`a scientifica di capire cosa
siano le armi all'UI e perch\'e sia stato scelto proprio l'uranio
impoverito per costruirle. Ad esempio, perch\'e non il ferro, il piombo o
leghe di altri metalli tradizionali? \`E possibile a chiunque, attraverso
una breve ricerca bibliografica, magari con l'ausilio della rete internet,
scoprire che le non meglio precisate armi all'UI sono sostanzialmente dei
proiettili anticarro (fig.~1): lunghe e sottili barre di uranio
238\footnote{Nella miscela isotopica di uranio-238 al 99.8\%, uranio-234
allo 0.001\% e uranio-235 allo 0.2\%, quindi {\it impoverita} dell'isotopo
fissibile 235, che nel minerale naturale si presenta con la percentuale
pi\`u alta dello 0.72\%.} con il compito di perforare le spesse e
resistenti corazze dei mezzi blindati. Questi proiettili non trasportano
nessun tipo di carica esplosiva: la loro azione sul bersaglio \`e
puramente meccanica (di sfondamento), e ovviamente non si innesca nessuna
reazione nucleare.  Il potere distruttivo di queste armi risiede quasi
esclusivamente nella loro capacit\`a di penetrazione: una volta che le
corazze vengono bucate, il materiale incandescente, prodotto dal processo
di penetrazione (essenzialmente metallo della corazza vaporizzato), si
proietta all'interno del mezzo bruciando tutto ci\`o che incontra. Ed \`e
facile allora che il mezzo blindato salti in aria a causa dell'esplosione
delle munizioni che esso stesso trasporta.


Sembra che l'efficacia dell'UI, come arma, stia essenzialmente nel fatto
che questo \`e un metallo ad alto peso specifico (circa $19.03$ $g/cm^3$).  
Nessuna differenza in qualit\`a e in quantit\`a sembra esserci invece nel
propellente usato per proiettare questi dardi rispetto a quello che viene
normalmente utilizzato per gli altri tipi di proiettili da cannone di
calibro compatibile.

E allora: in che modo una maggiore densit\`a del materiale di cui \`e
costituito un proiettile pu\`o incidere sulla sua capacit\`a di
penetrazione, quando la carica esplosiva di lancio, e quindi l'energia di
lancio, non viene aumentata? In fondo l'idea che il potere distruttivo di
un proiettile inerte\footnote{Cio\`e senza testata esplosiva.} dipenda
solo dalla sua energia cinetica appare pi\`u che ragionevole.

Il presente articolo \`e la risposta che siamo riusciti a dare a questo
interrogativo:  attraverso le due pi\`u importanti quantit\`a dinamiche
della fisica, cio\`e l'energia cinetica e l'impulso, \`e possibile
mostrare come l'uso di materiali ad alta densit\`a (ad esempio l'UI)
comporti un guadagno in capacit\`a di penetrazione rispetto ai materiali
tradizionali, a parit\`a di dimensioni fisiche del proiettile e di carica
esplosiva di lancio (cio\`e, a parit\`a di energia di lancio). Questo
punto \`e affrontato esplicitamente nelle sezioni {\bf 2} e {\bf 3}.

L'UI per\`o non \`e il solo materiale esistente a possedere un elevato
peso specifico.  Il tungsteno, ad esempio, con i suoi $19.3$ $g/cm^3$
pu\`o competere facilmente con l'UI e infatti presso le forze armate di
vari paesi sono in dotazione anche proiettili anticarro al tungsteno.  
L'UI ha avuto un maggiore successo poich\'e \`e anche ``a buon mercato'':
l'UI \`e il sottoprodotto del processo di arricchimento del combustibile
delle centrali nucleari e molti depositi sparsi in giro per il mondo ne
contengono grandi quantit\`a, l\`{\i} pronte ad essere utilizzate in
qualche modo.  Questo \`e brevemente l'argomento della quarta sezione.
Infine, nella sezione {\bf 5} mostrer\`o come una maggiore capacit\`a di
penetrazione si paghi con una diminuzione di gittata del proiettile.

\section{Modello di penetrazione}

Se $E_0$ \`e la frazione dell'energia dell'esplosivo che si trasforma in
energia cinetica del proiettile, allora

\begin{equation}
E_0=\frac{1}{2}Mv^2,
\label{eq1}
\end{equation}
dove $v$ \`e la veloct\`a del proiettile in uscita dal cannone e $M$ \`e la 
sua massa. Invertendo la (\ref{eq1}) si pu\`o ottenere la velocit\`a in 
funzione dell'energia e della massa

\begin{equation}
v=\sqrt{\frac{2E_0}{M}}.
\label{eq2}
\end{equation}

Consideriamo ora un modello molto semplificato di bersaglio e di
proiettile. Il proiettile \`e un cilindro di massa $M$, di superficie
frontale $S$ e lunghezza $\lambda$, dotato di energia cinetica $E_0$.
Stilizziamo invece il bersaglio come un semipiano occupato da particelle
di massa $m$ (arbitraria), con densit\`a numerica pari a $\sigma$
(fig.~2). Immaginiamo inoltre che ciascuna particella sia ``legata''
all'insieme da una energia di legame $E_l$. Se ad una singola particella
forniamo una quantit\`a di energia $E$ superiore a $E_l$, essa si distacca
dalla matrice e viaggia liberamente.


Il fenomeno reale di penetrazione di un proiettile in un bersaglio \`e
evidentemente molto pi\`u complesso e, possiamo dire, non ancora del tutto
compreso. Esso coinvolge fenomeni fisici che si verificano a ordini di
scala molto differenti: da complicati processi termodinamici (dimensione
atomica/molecolare) all'espulsione balistica di frammenti centimetrici.
Inoltre anche il proiettile si consuma, a volte fino a disintegrarsi
completamente. In questo studio, invece, assumiamo per semplicit\`a che il
proiettile rimanga integro durante tutto il processo. Assumiamo che
l'interazione tra proiettile e particelle del bersaglio sia una
interazione rigida: l'urto non modifica la massa e le propriet\`a di
entrambi. Supponiamo inoltre che gli urti avvengano in maniera
sequenziale, cio\`e il proiettile urta una particella del bersaglio alla
volta. Ciascun urto \`e non elastico nel senso che parte dell'energia
cinetica del proiettile (pari a $E_l$) deve essere assorbita affinch\'e la
particella di distacchi dalla matrice circostante.

Tenuto conto di questa non elasticit\`a del processo di urto, si possono
scrivere le leggi di conservazione dell'energia cinetica e della
quantit\`a di moto subito dopo che la particella si \`e distaccata e pu\`o
viaggiare liberamente

\begin{equation}
\left\{ \begin{array}{ll}
\sqrt{2M(E-E_l)}=M{\bar{v}}_1 + m{\bar{v}}_2  & {\rm cons.\quad 
impulso},\\
 E-E_l=\frac{1}{2}M{\bar{v}}^2_1 +\frac{1}{2}m{\bar{v}}^2_2 & {\rm 
cons.\quad energia},
\end{array}\right.
\label{eq4}
\end{equation}
dove ${\bar{v}_1}$ e ${\bar{v}_2}$ sono le velocit\`a del proiettile e della 
particella subito dopo l'urto.
L'energia a disposizione per il processo dinamico \`e $(E-E_l)$ poich\'e 
l'energia $E_l$ \`e ceduta dal proiettile alla particella $m$ per slegarla 
dalla matrice.

Il sistema di equazioni~(\ref{eq4}) pu\`o essere risolto agilmente
passando nel sistema di riferimento del centro di massa. I dettagli del
calcolo possono essere trovati in Landau e Lif\v sits~\cite{lali}.
Seguendo i calcoli descritti in~\cite{lali} si arriva ad ottenere per
$\bar{v}_2$, cio\`e la velocit\`a della particella dopo l'urto, la
seguente espressione

\begin{equation} 
\bar{v}_2=\frac{2M}{m+M}v_1, 
\label{eq5} 
\end{equation}
dove $v_1$ \`e la velocit\`a del proiettile subito dopo aver ceduto
l'energia di legame $E_l$ alla particella, e cio\`e
$v_1=\sqrt{\frac{2(E-E_l)}{M}}$.

L'energia cinetica della particella dopo l'urto \`e quindi

\begin{equation}
\bar{E}_2=\frac{1}{2}m\bar{v}^2_2=4\frac{m}{M}\frac{1}{\biggl(1+\frac{m}{M}\biggr)^2}(E
-E_l)\simeq 4\frac{m}{M}(E-E_l),
\label{eq6}
\end{equation}
poich\'e $m/M\ll 1$, cio\`e facciamo la ragionevole ipotesi che la massa dei 
frammenti (particelle) sia sempre estremamente pi\`u piccola della massa 
del proiettile.

Dunque, per ogni singolo urto la variazione totale di energia cinetica
subita dal proiettile \`e pari a

\begin{equation}
\Delta E=-\bar{E}_2-E_l\simeq -4\frac{m}{M}(E-E_l) -E_l,
\label{eq7}
\end{equation}
e per un numero opportunamente piccolo di incontri in sequenza 
${\rm d}n=S\sigma {\rm d}x$, dove ${\rm d}x$ \`e la penetrazione 
infinitesima del proiettile nel bersaglio, essa sar\`a pari a


\begin{equation}
{\rm d}E\simeq -S\sigma {\rm d}x\biggl(4\frac{m}{M}(E-E_l) +E_l\biggr).
\label{eq8}
\end{equation}

L'equazione (\ref{eq8}) pu\`o essere riscritta in forma di equazione 
differenziale di primo grado

\begin{equation}
\frac{{\rm d}E}{{\rm d}x}= -S\sigma \biggl(4\frac{m}{M}(E-E_l) 
+E_l\biggr).
\label{eq9}
\end{equation}

Se poniamo $F(x)=E(x)-E_l+\frac{M}{4m}E_l$, l'equazione (\ref{eq9})
diventa

\begin{equation}
\frac{{\rm d}F}{{\rm d}x}= -S\sigma 4\frac{m}{M}F(x),
\label{eq10}
\end{equation}
la cui soluzione generale \`e

\begin{equation}
F(x)= A\exp{\biggl[-S\sigma 4\frac{m}{M}x\biggr]}.
\label{eq11}
\end{equation}

Esplicitando $F(x)$ e ponendo come condizione iniziale $E(0)=E_0$ 
l'equazione (\ref{eq11}) diventa

\begin{equation}
E(x)= \biggl(E_0-E_l+\frac{M}{4m}E_l\biggr)
\exp{\biggl[-S\sigma 4\frac{m}{M}x\biggr]}+E_l-\frac{M}{4m}E_l.
\label{eq12}
\end{equation}

Quindi, la massima profondit\`a di penetrazione, $l_m$, si ha ponendo 
$E(l_m)=0$ nella (\ref{eq12}), cio\`e

\begin{table}[t]
{\small
\caption{\small\it Caratteristiche fisiche e dinamiche del proiettile.}
\begin{center}
\begin{tabular}{|c|c|c|}
\hline\hline\hline
$\lambda$ & $S$ & $E_0$ \\ 
  (cm)    &  (cm$^2$) & (J)    \\ \hline\hline
         &         &           \\
 $50$      & $7(^a)$     &  ${8.9\times 10^6}(^b)$        \\    
 & & \\ \hline\hline
\end{tabular}
\end{center}
Note: $(^a)$ questo valore corrisponde all'area di un cerchio di circa 
3 cm di diametro, $(^b)$ questa \`e l'energia cinetica di un corpo di 
circa 7 kg che viaggia a 1600 m/s.
}
\label{tabb}
\end{table}

\begin{equation}
0= \biggl(E_0-E_l+\frac{M}{4m}E_l\biggr)
\exp{\biggl[-S\sigma 4\frac{m}{M}l_m\biggr]}+E_l-\frac{M}{4m}E_l,
\label{eq13}
\end{equation}
e quindi, dopo qualche semplice passaggio algebrico,

\begin{equation}
l_m=\frac{M}{4S\sigma m}
\ln{\Biggl(\frac{E_0}{E_l\bigl(\frac{M}{4m}-1\bigr)}+1\Biggr)}.
\label{eq14}
\end{equation}

Ora siano $\rho_1$ e $\rho_2$ rispettivamente le densit\`a di massa del
proiettle e del materiale che costituisce il bersaglio. \`E facile vedere
che $\rho_2=\sigma m$. Tenendo conto del fatto che $M=\rho_1 \lambda S$ e
potendo ragionevolmente assumere che $\frac{M}{4m}\gg 1$, si ottiene per
$l_m$ la seguente espressione

\begin{equation}
l_m=\frac{\lambda\rho_1}{4\rho_2}\ln{\Biggl(\frac{E_0}{\frac{E_l}{m}
\frac{S\lambda\rho_1}{4}}+1\Biggr)}.
\label{eq15}
\end{equation}

La quantit\`a $\frac{E_l}{m}$ rappresenta l'energia di legame del
materiale del bersaglio per unit\`a di massa.  Cerchiamo ora di analizzare
il comportamento della~(\ref{eq15}) in funzione delle variabili che la
costituiscono.

\section{Esempio numerico: ferro, tungsteno e uranio im\-po\-ve\-ri\-to}

Immaginiamo che un proiettile con le dimensioni e l'energia cinetica
specificate nella tabella~1 colpisca una corazza. Mostreremo come la
quantit\`a $l_m$ cambi in funzione della sua densit\`a, $\rho_1$, e in
funzione del materiale che costituisce il bersaglio ($\rho_2$ e $E_l/m$).
Per comodit\`a poniamo $E_l/m$ uguale al calore di vaporizzazione del
metallo della corazza. Questa assunzione \`e probabilmente semplicistica
(anche per il fatto che per la costruzione di corazze vengono usate leghe
speciali e non metalli puri) ed \`e facile che i valori numerici assoluti
che si ottengono non siano compatibili con quelli di un esperimento reale.
Tuttavia lo scopo di questa nota \`e soprattutto quello di aiutare a
comprendere alcuni degli aspetti fisici pi\`u importanti che entrano in
gioco in un processo come questo.

La tabella~2 fornisce le densit\`a e i calori di vaporizzazione per i tre
metalli usati nel nostro esempio: ferro, tungsteno e uranio. I risultati
di questo esercizio sono esemplificati nella fig.~4. A prescindere dal
differente comportamento dei singoli materiali che costituiscono il
bersaglio, \`e evidente come l'aumento di $\rho_1$ implichi comunque un
aumento della profondit\`a di penetrazione $l_m$.

In sostanza, aumentare la densit\`a del proiettile, e quindi la sua massa,
a parit\`a di dimensioni rispetto ai proiettili tradizionali e di energia
cinetica iniziale $E_0$, riduce la frazione dell'energia $E_0$ che viene
dispersa sotto forma di energia cinetica dei frammenti e quindi aumenta la
quantit\`a di energia a disposizione per disgregare la materia del
bersaglio.

\begin{table}[t]
{\small
\caption{\small\it Densit\`a e calore di vaporizzazione di ferro, 
tungsteno e uranio.}
\begin{center}
\begin{tabular}{|c|c|c|}
\hline\hline\hline
Metallo & densit\`a & Calore di Vaporizzazione \\  
        &    (g/cm$^3$)    &  (KJ/g)     \\ \hline\hline  
        &           &   \\     
ferro (Fe) & 7.874 & 6.08806 \\  
tungsteno (W) & 19.30 & 4.48191 \\  
uranio (U) & 19.03  & 1.75188 \\  
& & \\ \hline\hline
\end{tabular}
\end{center}
}
\label{tab}
\end{table}

\section{Tungsteno e uranio impoverito}

Dal grafico~4 risulta evidente che le capacit\`a di penetrazione del
tungsteno puro e dell'uranio impoverito sono comparabili. Ma allora
perch\'e il pi\`u tossico uranio ha trovato una maggiore diffusione?  Le
motivazioni plausibili sono almeno due:

\begin{itemize}

\item[-] l'UI sembra essere pi\`u distruttivo poich\'e \`e anche un
metallo piroforico. Subito dopo essere penetrato all'interno della corazza
del carro, a contatto con l'aria, brucia spontaneamente. Ci\`o sembra
aumentare le capacit\`a incendiarie di queste armi.

\item[-] Ma soprattutto, l'UI, rispetto al tungsteno puro, \`e reperibile
in quantit\`a maggiori e a costi relativamente contenuti (es. scarti di
lavorazione nel processo di arricchimento del combustibile di centrali
nucleari). Bisogna tenere presente che l'UI \`e a tutti gli effetti una
scoria radioattiva da smaltire: nella discutibile ottica militare e
governativa, quale modo migliore per farlo?

\end{itemize}


\section{Gittata}

Infine, \`e interessante notare come i materiali ad alta densit\`a da una
parte aumentano il potere penetrante dei proiettili anticarro ma
dall'altra ne diminuiscono la gittata.  Questo fenomeno \`e ovviamente
connesso alla riduzione della velocit\`a di volo del proiettile, che si
pu\`o ricavare dall'equazione~(\ref{eq2}). L'espressione matematica per la
gittata balistica ideale di un corpo lanciato con velocit\`a $v$ e angolo
$\theta$ dal suolo \`e

\begin{equation} 
x_g=\frac{v^2}{g}\sin(2\theta), 
\label{eq17}
\end{equation} 
dove $g$ \`e l'accelerazione di gravit\`a. Quindi, usando la (\ref{eq2}) e 
avendo che $M~=~\rho~V$, il rapporto fra le gittate di due corpi di densit\`a 
diverse si esprime come

\begin{equation}
\frac{x_{g_1}}{x_{g_2}}=\frac{\rho_2}{\rho_1}.
\label{eq18}
\end{equation}

Quindi se $\rho_1$ \`e maggiore di $\rho_2$ la gittata $x_{g_1}$ diventa
minore della gittata $x_{g_2}$.

\section{Conclusioni}

In questo articolo si \`e sviluppato un semplice modello di penetrazione
di un proiettile in una corazza che permette di mostrare come, a parit\`a
di dimensioni fisiche (larghezza e lunghezza del proiettile) e di energia
cinetica, una maggiore densit\`a del materiale di cui il proiettile \`e
costruito contribuisca ad aumentare la sua profondit\`a di penetrazione.
Nella ragionevole ipotesi che l'energia cinetica del proiettile serva in
parte a disgregare la materia della corazza e in parte ad espellere i
frammenti prodotti (a fornire loro energia cinetica), una maggiore massa
del proiettile riduce la frazione di energia che si disperde sotto forma
di energia cinetica dei frammenti. Questo significa che una maggiore
quantit\`a di energia \`e a disposizione per il solo processo di
perforazione della corazza.

\`E per questo motivo, oltre al fatto che si tratta di un materiale
relativamente economico, che l'uranio impoverito tra gli altri materiali
competitori ha trovato ampia applicazione nella costruzione di proiettili
anticarro: esso infatti \`e uno dei pi\`u densi materiali esistenti in
natura. In questo articolo infine si \`e mostrato che una maggiore
densit\`a del materiale dei proiettili riduce significativamente la loro
gittata.

\section*{Ringraziamenti}

L'autore ringrazia il Prof. Paolo Paolicchi (Dip. Fisica, Universit\`a di
Pisa) per aver letto una prima versione dell'articolo ed aver contribuito
al suo miglioramento.  Un ringraziamento particolare va inoltre alla
Dott.ssa Barbara D'Abramo per aver reso l'italiano pi\`u fluido.

\newpage

\centerline{\bf Lista delle figure}

\vspace{1.5cm}

\noindent Fig.1 --- {\small Alcune munizioni anticarro a penetratore 
cinetico (sinistra) con spaccato (destra). Fonte~\cite{img}.}

\vspace{1.5cm}

\noindent Fig.2 --- {\small Schema del processo di impatto.}

\vspace{1.5cm}

\noindent Fig.3 --- {\small Proiettile anticarro all'UI in volo. Le parti
che si stanno distaccando dal dardo sono le strutture in alluminio che
servono ad adattare il proiettile al maggiore calibro del cannone di
lancio. Fonte~\cite{img}.}

\vspace{1.5cm}

\noindent Fig.4 --- {\small Profondit\`a di penetrazione, $l_m$, in
funzione della densit\`a del proiettile, $\rho_1$. Le tre curve si
riferiscono alle tre diverse composizioni della corazza; dall'alto verso
il basso: ferro, uranio e tungsteno. I punti delle curve contrassegnati
con le lettere Fe, U e W sono le profondit\`a raggiunte dai proiettili in
ferro, uranio e tungsteno nei tre diversi tipi di corazza.}

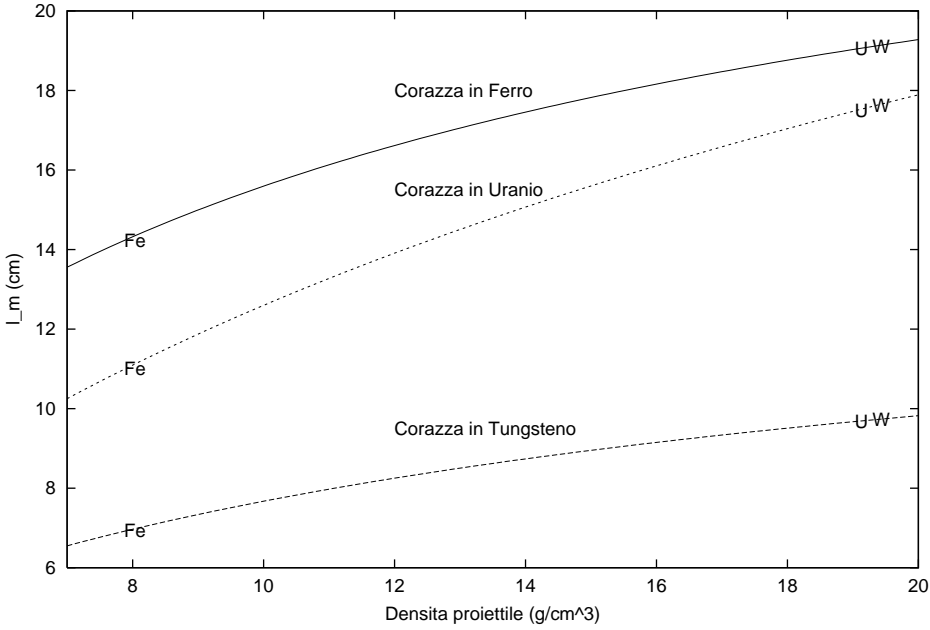
\end{document}